\documentclass[runningheads]{llncs}
\usepackage{graphicx}
\begin{document}
\title{Automatic segmentation of kidney and liver tumors in CT images}
%
%
\author{
Dina B. Efremova\inst{1},
Dmitry A. Konovalov\inst{1},
Thanongchai Siriapisith\inst{2},\\
Worapan Kusakunniran\inst{3}, 
Peter Haddawy\inst{3,4}
}
%
\authorrunning{D. Efremova et al.}
%

\institute{
    College of Science and Engineering, James Cook University, Townsville, 4811, Australia,    \email{dmitry.konovalov@jcu.edu.au}\\
\and
    Department Radiology, Faculty of Medicine, Siriraj Hospital, Mahidol University, Bangkok, 10700, Thailand
\and
    Faculty of Information and Communication Technology, Mahidol University, Nakhonpathom, 73170, Thailand
\and
    Bremen Spatial Cognition Center, University of Bremen, Germany
}
\maketitle              
\begin{abstract}
\boldmath
Automatic segmentation of hepatic lesions in computed tomography (CT) images is a challenging task to perform due to heterogeneous, diffusive shape of tumors and complex background. To address the problem more and more researchers rely on assistance of deep convolutional neural networks (CNN) with 2D or 3D type architecture that have proven to be effective in a wide range of computer vision tasks, including medical image processing. In this technical report, we carry out research focused on more careful approach to the process of learning rather than on complex architecture of the CNN. We have chosen MICCAI 2017 LiTS dataset for training process and the public 3DIRCADb dataset for validation of our method. The proposed algorithm 
reached DICE score 78.8\% on the 3DIRCADb dataset. The described method was then applied to 
the 2019 Kidney Tumor Segmentation (KiTS-2019) challenge, where our single submission achieved 96.38\% for kidney and 67.38\% for tumor Dice scores.

\end{abstract}
\section{Introduction}

Hepatic tumor may cause a serious threat to human health and lives. To prevent and monitor liver diseases it is important to provide accurate segmentation of abnormal tissues in the organ. Although liver segmenting task has achieved good results thanks to CNN, localization of liver tumors is still a demanding problem and has some room for improvement.

CT is generally used image modality by radiologists and oncologists for liver tumor evaluation, but sometimes CT scans have noise in them due to reduction of the CT radiation dose, which is always a trade-off between image quality and health risks for patients. Other main issues in the segmentation task are the large scale of spatial and structural variability, low contrast between liver and tumor tissues, high variation in size, shape and number of lesions, even the similarity of nearby organs.

To date, segmentation of biomedical and medical images is an active research area.
In this work we adopted the delineation between methods based on how autonomous they were actually detecting liver and/or liver tumor.
According to this, algorithms for segmentation fall into groups of semi-automatic and fully-automatic techniques.
The methods for segmentation of CT images are reviewed next.

\subsection{Related semi-automatic methods}

The related studies based on semi-automatic methods are reviewed here.
In all semi-automatic methods, a qualified radiologist must first locate liver and/or liver tumors manually by selecting a bounding box or other area selection.

In early attempts to perform segmentation might be used such technique as graph cut \cite{Boykov1999}, when computer vision approach could not operate very deep and extensive networks, so it highly relied on complex mathematical base.

In 2005, Liu et al. \cite{Liu2005} developed a method for segmentation of the liver contour, where a Canny edge detector was used together with a snake algorithm and a gradient vector flow (GVF) field as its external force. 
The method achieved a median 5.3\% error by segmentation volume on 551 2D liver $512 \times 512$ images.
This category of methods based on the local pixel intensity and/or gradients were actively 
explored \cite{Li2012} and demonstrated reasonable results in liver tumor segmentation even in low contrast CT images.

Siriapisith et al. have proposed a 2D segmentation method \cite{Siriapisith2018} that applies the concept of variable neighborhood search by iteratively alternating search through intensity and gradient spaces. They have claimed to achieve the segmentation performance with a DSC of $84.48\pm5.84$\% and $76.93\pm8.24$\% for large and small liver tumor respectively.

A texture-specific BoVW method \cite{Xu2018} for the  retrieval of focal liver lesions has been introduced by Xu et al. The bag of visual words (BoVW) model is meant for feature representation that can integrate various handcrafted features like intensity, texture, and spatial information and thus is able to effectively characterize various liver tumors.

Zheng et al. \cite{Zheng2018} have presented the method which combines hybrid algorithms such as a unified level set method (LSM) coupled with hidden Markov random field and expectation-maximization (HMRF-EM). The proposed LSM approach incorporates both region information and edge information to evolve the contour and it is more resistant to edge leakage than the single-information driven LSMs.

Interaction is required for the mentioned above liver and liver tumor segmentation methods, that fact restricts using those frameworks with well-qualified specialists.

\subsection{Related fully-automatic methods}
The major recent breakthrough in the field of semantic segmentation is widely attributed to the general-domain fully convolutional neural networks (FCNs) of \cite{FCN15}, \cite{FCN17} and biomedical-domain U-net of \cite{unet2015}. 
The term {\em semantic} means that each pixel is assigned a label or a class of objects during training and prediction phases. 
For example, in this context, each image pixel could be assigned one of three labels: {\em liver}, {\em tumor}, {\em other}.
Therefore the semantic segmentation could be conveniently defined as a per-pixel classification problem.

Since the CNN methods have been developing at accelerating rate, it is revealing to note their publication years.

In 2015, for liver tumor segmentation, Li et al. \cite{Li2015} achieved the Dice similarity coefficient \cite{Dice1945MeasuresOT} (DICE, see Eq.~\ref{eq:DICE}) \mbox{of 80\%} and the area under the Precision-Recall curve of 0.9556, where the results were the averages of 30 leave-one-out cross validations on 30 CT images. 
Most notably it was clearly demonstrated that their CNNs outperformed other traditional machine learning methods such as AdaBoost, random forest (RF), and support vector machine (SVM).
Gaussian smoothing filter was used as prepossessing, and the images were downsized by a factor of 2. 
Four different CNN architectures (with 6 or 7 layers) were tested, where their input shapes ranged between $13 \times 13$ and $19 \times 19$ grayscale image patches. 

The CNN approach in \cite{Li2015}
did not follow the currently most widely used segmentation CNN architectures \cite{FCN15}, \cite{FCN17}, \cite{unet2015}.
The fully convolutional neural networks (FCNs): in VGG-FCN \cite{FCN15} and U-Net\cite{unet2015} CNNs, there are two distinct sections: {\em encoder} and {\em decoder}.
Encoder layers or even a complete classification CNN (for example, VGG16 \cite{VGG16} in \cite{FCN15}) downsize the input image by up to 32 times while the image features are extracted. 
Then the decoder layers reconstruct the original input shape with the required number of segmentation layers or channels.
In \cite{Li2015}, only the encoder-type CNN was used by centering it at each input pixel.

In 2016, Dou et al. \cite{Dou2016} developed a 3D version of the VGG-FCN \cite{FCN15} architecture with deep supervision to hidden layers, so-called 3D deeply supervised network (3D DSN), which could accelerate the optimization convergence rate and improve the prediction accuracy.
Additionally, 3D DSN generated the high-quality score map that helped to make contour refinement with a fully connected conditional random field (CRF) to obtain refined segmentation results.

Lu et al. \cite{Lu2017} proposed method for liver segmentation consisted of two steps: first, using 3D CNNs to detect liver and make probabilistic segmentation, and second, to refine accuracy of initial segmentation with graph cut and the previously learned probability map.
The suggested approach was validated on 3DIRCADb dataset and reached 9.36\% and 0.97\% for volume overlap error (VOE, see Eq.~\ref{eq:VOE_RVD}) and relative volume difference (RVD, see Eq.~\ref{eq:VOE_RVD}) respectively.

In 2017, Christ et al. \cite{Christ2017AutomaticLA} trained two cascaded UNet-type \cite{unet2015} FCNs. 
The first FCN segmented a liver out of the rest of the inner body tissues.
Then the second FCN segmented lesions from the output ROIs (regions-of-interest) of the first FCN.
Dense 3D CRF was used as the post-processing to refine the FCN predictions.
DICE over  94\% was achieved on the 15 hepatic tumor volumes from the  abdominal CT dataset 3DIRCADb \cite{3DIRCADb} for liver segmentation and 56\% for lesions.

Sun et al. \cite{Sun2017AutomaticSO} used a segmentation CNN conceptually similar to the FCN architecture of \cite{FCN15}, where the \mbox{AlexNet \cite{AlexNet}} CNN was used as the encoder.
On the 3DIRCADb dataset, Sun et al. reported the VOE of $15.6 \pm 4.3\%$.
In addition to the publicly available 3DIRCADb dataset, the private JDRD dataset was labeled by two radiologists at The First Hospital of Jilin University. 
The unique feature of the JDRD dataset was its three CECT (contrast-enhanced computed tomography) per-slice images taken at three blood flow phases: arterial (ART), portal venous (PV), and delayed (DL) phase at the same lesion locations.
When all three grayscale phase-specific images where combined into three-channel images, the studied multi-channel segmentation CNN (MC-FCN) improved \mbox{VOE = 8.1 $\pm$ 4.5\%}.

In 2018, Li et al. \cite{Li2018-DenseUNet} developed a hybrid densely connected UNet (H-DenseUNet), which  consists of a 2D DenseUNet and a 3D counterpart.
H-DenseUNet worked in an end-to-end manner, where the intra-slice representations and inter-slice features can be jointly optimized through a hybrid feature fusion (HFF) layer for accurate liver and lesion segmentation.
H-DenseUNet was trained on MICCAI 2017 Liver Tumor Segmentation (LiTS) dataset \cite{lits2017} and validated on the 3DIRCADb dataset achieving the liver \mbox{DICE = 98.2\%} and tumor \mbox{DICE = 93.7\%}.
Worth noting that they also conducted experiments exclusively on 3DIRCADb dataset through cross-validation and achieved the liver segmentation \mbox{DICE = 94.7\%} and tumor \mbox{DICE = 65\%}.

Some recent papers have been dedicated to attention mechanism, like \cite{Jin2018RAUNetAH} and \cite{Jiang2019AHCNetAA}.

In 2018, Jin et al. \cite{Jin2018RAUNetAH} used a 3D hybrid residual attention-aware segmentation method, called RA-UNet, in their experiments.
Attention modules were stacked so that the attention-aware features could change adaptively as the network went "very deep" due to residual learning.
The model was trained on MICCAI 2017 LiTS dataset and validated on 3DIRCADb with DICE 97.7\% and 83\% for liver and lesion segmentation respectively.

In 2019, Jiang et. al \cite{Jiang2019AHCNetAA} proposed a 3D FCN structure, composed of multiple Attention Hybrid Connection Blocks (AHCBlocks) densely connected with both long and short skip connections and soft self-attention modules. Same training process with LiTS and 3DIRCADb datasets estimated DICE 95.9\% and 73.4\% for liver and tumor segmentation accordingly.

At the moment a large number of solutions have been proposed for liver tumor segmentation from CT images. 
The fully-automatic methods have received major attention recent years, because it is meant to lift burden of segmentation from human experts and exclude human bias and mistakes.

\begin{figure}
\includegraphics[width=\textwidth]{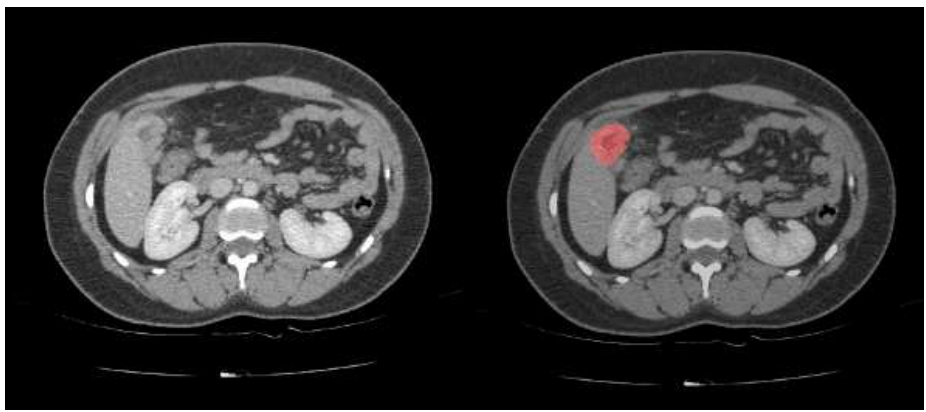}
\caption{An example of contrast-enhanced CT scan showing the difficult case of lesion segmentation when edges of the tumor barely distinguishable in the original image (on the left). The red region denotes the ground truth mask of liver lesion (on the right).}\label{fig1}
\end{figure}

\section{Methods}

\subsection{Datasets}

\subsubsection{Liver}

For this research two datasets have been used:
\mbox{MICCAI 2017} Liver Tumor Segmentation (LiTS) Challenge \cite{lits2017} and 3DIRCADb \cite{3DIRCADb} (3D Image Reconstruction for Comparison of Algorithm Database).

The LiTS dataset contains of 201 contrast-enhanced 3D abdominal CT volumes with different types of tumor contrast levels, abnormalities in tissues size and varying amount of lesions. We have used this dataset for training our model.

The 3DIRCAD dataset includes 20 venous phase enhanced CT volumes from various European hospitals with different CT scanners involving 120 liver tumors of different sizes. We have evaluated our method on this dataset.

Expert radiologists have manually outlined liver tumor contours for all images on a slice-by-slice basis in order to determine the ground truth. The 3Dircadb dataset is segmented by a single radiologist, while the LiTS dataset is created in collaboration with seven hospitals and research institutions and manually reviewed by independent three radiologists.

Because of imbalanced classes, liver tumor areas are significantly less than background, we have applied data augmentation to the training dataset. Such techniques as elastic transformation, shifting, scaling, and rotating have been used.

\subsubsection{Kidney}
The 2019 Kidney Tumor Segmentation (KiTS) Challenge \cite{KiTS19} training dataset contained 210 different patients. 
The KiTS challenge required automatic segmentation of 90 test patients for which the ground truth segmentations were not released before the submission due date (29th of July, 2019). 

\subsection{Semantic Segmentation of Images}

\begin{figure}
\includegraphics[width=\textwidth]{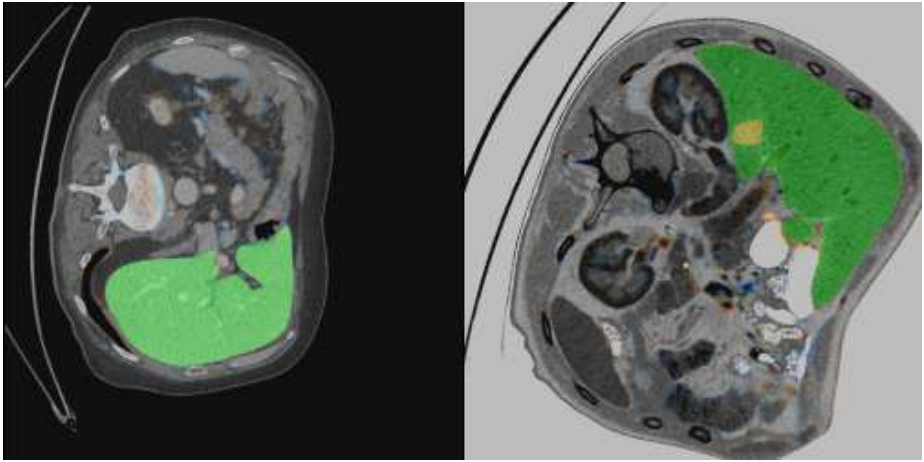}
\caption{An example of the method segmentation: liver (green) and liver tumor (yellow) segmentation.}
\label{fig2}
\end{figure}

For the purpose of this paper we have selected a variation of U-Net \cite{unet2015}, LinkNet-34 \cite{Chaurasia2017LinkNetEE}, where ResNet-34 \cite{He2016DeepRL} has been used as the feature encoder and PyTorch implementation was from \cite{Shvets2018AutomaticIS}. LinkNet-34 has a reasonable number of parameters and a good balance between running time and accuracy.

One of the problems of deep learning with the CNN is that the learning phase, where the network undergoes training process from scratch, can be very time-consuming and may need a very large set of images.
A simple yet effective transfer learning strategy is introduced to overcome several problems at once.
First of all, pre-trained weights are already learnt to recognize patterns in images, the network needs less time to converge to a new solution, usually a better solution than in case of training from scratch.
Transfer learning prevents or at least mitigates over-fitting problem.

In our method, we have reused the ImageNet-trained ResNet-34 encoder without freezing the weights during training process due to more advanced segmentation CNN
(compared to FCN-8s).
However, to assist in more effective way of using the pre-trained network, the learning rate has been reduced by factor of 10 when applied to the encoder, whereas to the randomly initialized LinkNet-34 decoder
layers, the learning rate without any change has been applied.

The input layer of the LinkNet-34 model has been modified, instead of original \mbox{3-channel} RGB colour we have conducted experiments with single channel and so-called 2.5D architecture \cite{Han2017AutomaticLL}.
Single channel is applicable due to that all CT scans are grayscale and have only one colour channel.
2.5D architecture proposed by Han \cite{Han2017AutomaticLL} is a 2D deep CNN which takes a stack of adjacent slices from the volumetric images as input and produces the segmentation map corresponding to the center slice.

As for the loss function, the binary cross entropy with negative DICE coefficient (see Eq.~\ref{eq:loss}) has been used,

\begin{equation} 
\label{eq:loss}
loss(y, \hat{y}) = bc(y, \hat{y}) -  log(dice(y, \hat{y})),
\end{equation}

where $y$ is a target mask, $\hat{y}$ is the corresponding LinkNet-34 output, $bc(y, \hat{y})$ is the binary cross entropy, $dice(y, \hat{y})$ is the DICE coefficient.

Training process has been performed on the LiTS dataset consisted of 131 patients with 58,638 image-mask pairs, whereas validation on 3DIRCADb dataset has been done with two different approaches: tested on tumors larger than 100-pixel area and on tumors of any size.

\section{Results}

\begin{table}
\caption{Comparison of liver and liver tumor segmentation results on 3DIRCAB dataset.}
\begin{center}
\begin{tabular}{|l|c|c|c|l|}
\hline
Method  & VOE(\%) & RVD(\%) & DICE(\%) & Type\\
\hline
Li et al. \cite{Li2018-DenseUNet} 
& 3.57 $\pm$ 1.66 & 0.01 $\pm$ 0.02 & 98.2 $\pm$ 1 & liver \\
H-DenseUNet
& 11.68 $\pm$ 4.33 & -0.01 $\pm$ 0.05 & 93.7 $\pm$ 2 & tumor \\
\hline
Deng et al. \cite{Deng2019DynamicRO}
&  &  & & \\
3D CNN
& 26.93 $\pm$ 8.51 & 6.55 $\pm$ 14.91 & 85 $\pm$ 6 & tumor\\
\hline
Jin et al. \cite{Jin2018RAUNetAH}
& 4.5 & -0.1 & 97.7 & liver\\
RA-UNet
&  &  & 83 & tumor \\
\hline
Huang et al. \cite{Huang2018RobustEF}
&  &  &  & \\
semi-automatic
& 27.05 $\pm$ 9.19 & 4.23 $\pm$ 19.28 & 84 $\pm$ 7 & tumor\\
\hline
Jiang et al. \cite{Jiang2019AHCNetAA}
&  &  & 95.9 & liver \\
AHCNet
& 1.35 & 0.13 & 73.4 & tumor \\
\hline
our
&  &  & 96.2 & liver \\
LinkNet-34 \cite{Chaurasia2017LinkNetEE}
&  &  & 78.8 & tumor \\
\hline
\end{tabular}
\end{center}
\label{Tab:tumor}
\end{table}

\subsection{Metrics}

To evaluate performance of the segmentation task different metrics are applied, although we have focused on widely used ones that are usually utilized for liver and liver tumor segmentation such as 
the mean ratios of volume overlap error (VOE),
relative volume difference (RVD),

\begin{equation} 
\label{eq:VOE_RVD}
\mbox{VOE} =   \left [ 1 - \frac{|A \cap B|}{|A \cup B|} \right ], \ \ \ 
\mbox{RVD} =   \frac{|A| - |B|}{|B|},  
\end{equation}

and Dice similarity coefficient (DICE) \cite{Li2015}, \cite{Sun2017AutomaticSO}, \cite{Siriapisith2018-outer-wall}.
\begin{equation} 
\label{eq:DICE}
\mbox{DICE} =  \frac{ 2 \times |A \cap B|}{|A| + |B|},
\end{equation}
where $A$ it the segmentation result and $B$ is the ground truth.

\subsection{Comparison to other methods}

As mentioned before, we have trained the model on the LiTS dataset and evaluated it on the 3DIRCADb dataset. For comparison, we have chosen papers with the similar approach, see Table~\ref{Tab:tumor}.

Some techniques have been more useful than others for the segmentation task. For example, data augmentation has had positive contribution to the accuracy of the method, while utilizing widely used 2.5D approach has not improved any DICE metrics.
The approach with different learning rates has allowed to converge the model more quickly and to a better solution.

\section{Conclusion}

Our goal was to research a different approach to the segmentation task and show that the training pipeline could matter as much as the CNN architecture.
While using of advanced CNN models may be constrained by hardware and the complexity of their implementation, a customized training
pipeline could achieve competitive baseline
results with relatively simple CNNs in fraction of time 
what would normally required for more complex CNNs. 
We deliberately selected an
off-the-shelf CNN (LinkNet-34), which was not the state-of-the-art network. Consistently applying different kinds of techniques, we have reached competitive results and outperformed at least one compound CNN \cite{Jiang2019AHCNetAA} for liver and liver tumor segmentations.

The proposed method was applied to
the 2019 Kidney Tumor Segmentation Challenge \cite{KiTS19}, and the
corresponding results were submitted for evaluation achieving the 38th place out of 106 submissions, where our Dice scores were 0.9638 (kidney), 0.6738 (tumor), and 0.8188 (composite, i.e. mean of kidney and tumor scores).

%
%
%
%

\end{document}